\documentclass[aps,amsmath,amssymb,twocolumn,twoside,floatfix,preprintnumbers,nobibnotes,nofootinbib]{revtex4}

\usepackage{graphicx}
\usepackage{enumerate}

\newcommand{\bea}{\begin{eqnarray}}
\newcommand{\beq}{\begin{equation}}
\newcommand{\eea}{\end{eqnarray}}
\newcommand{\eeq}{\end{equation}}

\newcommand{\tgb}{ t_\beta }

\newcommand{\be}{\begin{equation}}
\newcommand{\ee}{\end{equation}}

\newcommand{\ba}{\begin{array}}
\newcommand{\ea}{\end{array}}

\def\R1{\varepsilon_1}
\def\E8{\varepsilon_8}

\newcommand{\bd}{\begin{displaymath}}
\newcommand{\ed}{\end{displaymath}}

\newcommand{\bi}{\begin{itemize}}
\newcommand{\ei}{\end{itemize}}

\newcommand{\SU}[1]{\ensuremath{\mathrm{SU}(#1)}}

\def\npb#1#2#3{    {\it Nucl. Phys. }{\bf B #1} (#2) #3}

\begin{document}
\preprint{TUM-HEP-725/09}

\title{The SUSY CP Problem and the MFV Principle}

\author{Paride Paradisi}
\author{David M. Straub}
\affiliation{Physik-Department, Technische Universit\"at M\"unchen,
85748 Garching, Germany}

\begin{abstract}

We address the SUSY CP problem in the framework of Minimal Flavor Violation (MFV), where the
SUSY flavor problem finds a natural solution. By contrast, the MFV principle does not solve the
SUSY CP problem as it allows for the presence of new {\it flavor blind} CP-violating phases.
Then, we generalize the MFV ansatz accounting for a natural solution of it. The phenomenological
implications of the generalized MFV ansatz are explored for MFV scenarios defined both at the
electroweak (EW) and at the GUT scales.

\end{abstract}

\maketitle

\section{Introduction}\label{sec:intro}
Supersymmetric (SUSY) extensions of the Standard Model (SM) are broadly considered as the
most motivated and promising New Physics (NP) theories beyond the SM. The solution of the
gauge hierarchy problem, the gauge coupling unification and the possibility of having a
natural cold dark matter candidate, constitute the most convincing arguments in favor of SUSY.

On the other hand, a generic SUSY scenario provides many (dangerous) new sources of flavor
and CP violation, hence, large non-standard effects in flavor processes would be typically
expected.

However, the SM has been very successfully tested by low-energy flavor observables both
from the kaon and $B_d$ sectors.

In particular, the two $B$ factories have established that $B_d$ flavor and CP violating
processes are well described by the SM up to an accuracy of the $(10-20)\%$ level~\cite{hfag}.

This immediately implies a tension between the solution of the hierarchy problem, calling
for a NP scale below the TeV, and the explanation of the Flavor Physics data requiring
a multi-TeV NP scale if the new flavor-violating couplings are generic.

An elegant way to simultaneously solve the above problems is provided by the Minimal Flavor
Violation (MFV) hypothesis~\cite{MFV,MFV_gen}, where flavor and CP violation are assumed to
be entirely described by the CKM matrix even in theories beyond the SM.

However, the MFV principle does not provide in itself any restriction to the presence of new CP-violating phases, hence, the assumption that the CKM phase provides the only source for CP violation (CPV) even
in NP theories satisfying the MFV principle seems to be not general and thus  a restrictive assumption~\cite{colangelo_MFV,smithEDM} (see also~\cite{EllisCP,ABP,kaganMFV,feldmannNEW}).

In this context, we analyze the most general SUSY scenario, compatible with the MFV principle,
allowing for the presence of new CP violating sources.

In general, a MFV MSSM suffers from the same SUSY CP problem as the ordinary MSSM. In fact, the
symmetry principle of the MFV does not forbid the presence of the dangerous {\it flavor blind}
CP violating sources such as the $\mu$ parameter in the Higgs potential or the trilinear scalar
couplings $A_I$.
When such phases assume natural $\mathcal O(1)$ values and if the SUSY scale is not far from the
EW scale, the bounds on the EDMs of the electron and neutron are violated by orders of magnitude:
this is the so-called SUSY CP problem.

Either an extra assumption or a mechanism accounting for a natural suppression of these CPV phases
are desirable.

In this work, we assume a {\it flavor blindness} for the soft sector, i.e. universality of the soft
masses and proportionality of the trilinear terms to the Yukawas, when SUSY is broken. In this limit,
we also assume CP conservation and we allow for the breaking of CP only through the MFV compatible
terms breaking the {\it flavor blindness}.

That is, CP is preserved by the sector responsible for SUSY breaking, while it is broken in the
flavor sector.

The generalized MFV scenario naturally solves the SUSY CP problem while leading to specific and
testable predictions in low energy CP violating processes.

\section{CP violation in SUSY MFV scenarios}
\label{CPV_MFV}
The hypothesis of MFV states that the SM Yukawa matrices are the only source of flavor breaking,
even in NP theories beyond the SM~\cite{MFV,MFV_gen}. The MFV ansatz offers a natural way to avoid
unobserved large effects in flavor physics and it relies on the observation that, for vanishing
Yukawa couplings, the SM enjoys an enhanced global symmetry
\begin{equation}\label{eq:Gflavor}
 G_\mathrm{f}~=~
\SU3_u \times \SU3_d \times \SU3_Q \times \SU3_e \times \SU3_L\,.
\end{equation}
The SM Yukawa couplings are formally invariant under $G_\mathrm{f}$ if the Yukawa matrices are
promoted to spurions transforming in a suitable way under $G_\mathrm{f}$.
NP models are then of the MFV type if they are formally invariant under $G_\mathrm{f}$, when
treating the SM Yukawa couplings as spurions.

In the MSSM with conserved $R$-parity, the most general expressions for the low-energy soft-breaking
terms compatible with the MFV principle and relevant for our analysis read~\cite{colangelo_MFV}
\bea
\mathbf{m}_{Q}^{2} &=& m_{Q}^{2}
\bigg[
\mathbf{1}+r_{1}\mathbf{Y}_{u}^{\dagger}\mathbf{Y}_{u}+r_{2}\mathbf{Y}_{d}^{\dagger}\mathbf{Y}_{d}+
\nonumber\\
&+&
(c_{1} \mathbf{Y}_{d}^{\dagger}\mathbf{Y}_{d}\mathbf{Y}_{u}^{\dagger}\mathbf{Y}_{u}+\mathrm{h.c.})
\bigg]\;,
\label{MFV:mQ}
\eea
\bea
\mathbf{m}_{D}^{2} &=& m_{D}^{2}
\bigg[\mathbf{1}+\mathbf{Y}_{d}
\bigg(r_{3} +r_{4}\mathbf{Y}_{u}^{\dagger}\mathbf{Y}_{u}+r_{5}\mathbf{Y}_{d}^{\dagger}\mathbf{Y}_{d}
\nonumber\\
&+&
(c_{2}\mathbf{Y}_{d}^{\dagger}\mathbf{Y}_{d}\mathbf{Y}_{u}^{\dagger}\mathbf{Y}_{u}+\mathrm{h.c.})
\bigg)
\mathbf{Y}_{d}^{\dagger}
\bigg]\;,
\label{MFV:mD}
\eea
\bea
\mathbf{A}^{U} &=&
A_{U}\mathbf{Y}_{u}\bigg( \mathbf{1} + c_{3}\mathbf{Y}_{d}^{\dagger}\mathbf{Y}_{d}+
c_{4}\mathbf{Y}_{u}^{\dagger}\mathbf{Y}_{u} +
\nonumber\\
&+&
c_{5}\mathbf{Y}_{d}^{\dagger}\mathbf{Y}_{d}\mathbf{Y}_{u}^{\dagger}\mathbf{Y}_{u}+
c_{6}\mathbf{Y}_{u}^{\dagger}\mathbf{Y}_{u}\mathbf{Y}_{d}^{\dagger}\mathbf{Y}_{d}
\bigg)\;,
\label{MFV:AU}
\eea
\bea
\mathbf{A}^{D} &=&
A_{D}\mathbf{Y}_{d}\bigg(\mathbf{1}+c_{7}\mathbf{Y}_{u}^{\dagger}\mathbf{Y}_{u}+
c_{8}\mathbf{Y}_{d}^{\dagger}\mathbf{Y}_{d}+
\nonumber\\
&+&
c_{9}\mathbf{Y}_{d}^{\dagger}\mathbf{Y}_{d}\mathbf{Y}_{u}^{\dagger}\mathbf{Y}_{u}+
c_{10}\mathbf{Y}_{u}^{\dagger}\mathbf{Y}_{u}\mathbf{Y}_{d}^{\dagger}\mathbf{Y}_{d}
\bigg)\;,
\label{MFV:AD}
\eea
where $m_{Q}$, $m_{D}$, $A_U$ and $A_D$ set the mass scale of the soft terms, while $r_i$
and $c_i$ are unknown, order one, numerical coefficients.

Notice that, in the above expansions, the SM Yukawa couplings are not assumed to be the only
source of CPV as done instead in~\cite{MFV_gen}. In particular, while all the $r_i$ parameters
must be real, as the squark mass matrices are hermitian, the $c_i$ parameters are generally complex~\cite{colangelo_MFV}.

As in the ordinary MSSM, flavor conserving CP violating sources such as the $\mu$ parameter in
the Higgs potential or the trilinear scalar couplings $A_I$ are unavoidable also in SUSY MFV frameworks, as they are not forbidden by the symmetry principle of the MFV~\cite{colangelo_MFV}.

Physics observables will then depend only on the phases of the combinations $M_i\mu$, $A_I\mu$
and $A_I^{*}M_i$~\cite{pospelov} and it is always possible to choose a basis where only the $\mu$
and $A_I$ parameters remain complex~\footnote{To be precise, such a statement is valid as long as
the gaugino masses are universal at some scale. Even in this last case, two loop effects driven
by a complex stop trilinear $A_t$ generate an imaginary component for $M_i$~\cite{oliveCP} that
we systematically take into account in our numerical analysis.}.

These CP violating phases generally lead to too large effects for the electron and neutron EDMs,
which are induced already at the one loop level through the virtual exchange of gauginos and
sfermions of the first generation.

In particular, the current experimental bounds on the electron~\cite{expedme} and
neutron~\cite{expedm} EDMs imply that
\begin{eqnarray}
|\sin\phi_\mu| &\lesssim&
 10^{-3}\left(\frac{m_{{\rm SUSY}}}{300~{\rm GeV}}\right)^{2}\!\!
 \left({\frac{10}{t_{\beta}}}\right)\,,
\nonumber\\
|\sin\phi_A| &\lesssim&
 10^{-2}\left(\frac{m_{{\rm SUSY}}}{300~{\rm GeV}}\right)^{2}\,,
\nonumber
 \label{nedm}
\end{eqnarray}
if we impose the bounds on $\phi_\mu$ and $\phi_A$ separately. In Eq.~(\ref{nedm}),
$t_{\beta}=\tan\beta$ and a common SUSY mass $m_{{\rm SUSY}}$ has been assumed.

The naturalness problem of so small CP-violating phases, provided a SUSY scale of the order of
the EW scale, is commonly referred to as the SUSY CP problem. Hence, either an extra assumption
or a mechanism accounting for such a strong suppression in a natural way are desirable.
\section{A generalized MFV ansatz and the SUSY CP problem}
\label{CP_MFV_ANSATZ}
The SUSY CP problem is automatically solved in the MFV framework of D'Ambrosio et al.~\cite{MFV_gen},
as they assume the extreme situation where the SM Yukawa couplings are the only source of CPV.

However, the MFV symmetry principle allows for the presence of new CPV phases, in particular
of {\it flavor blind} phases that represent the main source of the SUSY CP problem.

Instead of following the approch of D'Ambrosio et al.~\cite{MFV_gen}, we first observe that
the assumption of the {\it flavor blindness} corresponds to setting all the $r_i$ and $c_i$
coefficients to zero.

In this limit, we assume CP conservation and we allow for the breaking of CP only through the
terms breaking the {\it flavor blindness}.

In this way, $A_{U}$, $A_{D}$, the gaugino masses and the $\mu$ term turn out to be real
at the scale where the MFV holds while the leading imaginary components of the $A$
terms, induced by the complex parameters $c_i$, have a cubic scaling with the Yukawas.

Notice that, after the infinite sum of MFV-compatible terms for Eqs.~(\ref{MFV:mQ})-(\ref{MFV:AD})
is taken into account, the generation of CP-violating phases for $A_U$ and $A_D$ is
unavoidable~\cite{colangelo_MFV,smithEDM,smith_private}. However, we have checked that
these phases are at most of order $\sim y^2_c \sim 10^{-4}$, hence, safely neglegible.

If we now deal with a low scale MFV scenario, the one loop contributions to the electron and
neutron EDMs, that depend on the first generation $A$ terms, are proportional to the cube of
light fermion masses, hence safely under control even for order one CPV phases of the $c_i$
parameters.
As a result, the generalized MFV ansatz applied to a low scale SUSY MFV scenario can completely
cure the SUSY CP problem.

The situaton can drastically change if we define a SUSY MFV scenario at the GUT scale.

In this last case, RGE effects stemming from trilinears of the third generation will unavoidably
generate complex trilinears for light generations and a complex $\mu$ term at the low scale.
As a result, the EDMs will receive both one and two loop contributions and the SUSY CP problem
might reappear. However, as we will discuss in detail later, if CPV arises only from terms
breaking the {\it flavor blindness}, it will be still possible to account for the SUSY CP
problem in natural ways.


But how natural is the assumption for the origin of CP breaking in the generalized MFV scenarios?

As an attempt to address this question, we make a comparison between the generalized MFV scenario
and SUSY flavor models.

In fact, one could envisage the possibility that the peculiar flavor structure of the soft-sector
dictated by the MFV principle might be the remnant of an underlying flavor symmetry holding at 
some high energy scale.

Supersymmetric models with {\it abelian}~\cite{abelian,nir_rattazzi} and
{\it non-abelian}~\cite{nonabelian,ross} flavor symmetries have been extensively discussed in
the literature. They are based on the Frogatt-Nielsen~\cite{FN} mechanism where the flavor
symmetries are spontaneously broken by (generally complex) vacuum expectation values of some ``flavon'' fields $\Phi$ and the hierarchical patterns in the fermion mass matrices can then
be explained by suppression factors $\left(\langle\Phi\rangle/M\right)^n$, where $M$ is the
scale of integrated out physics and the power $n$ depends on the horizontal group charges of
the fermion, Higgs and flavon fields.

Then, such flavor symmetries, while being at the origin of the pattern of fermion masses and mixings, relate, at the same time, the flavor structure of fermion and sfermion mass matrices.

However, the CP violating effects to the EDMs driven by the {\it flavor blind} phases
$\phi_A,\phi_{\mu}$ are, in general, not constrained at all by the flavor symmetry and
an additional assumption is required.

The usual assumption employed by SUSY flavor models is that CP is a symmetry of the theory
that is spontaneously broken only in the flavor sector as a result of the flavor symmetry breaking~\cite{nir_rattazzi,ross}.

Hence, we believe that the assumption we made on the origin of CPV in MFV scenarios is reminiscent of the usual approach followed in SUSY flavor models.

In the light of these considerations, we proceed now to analyze the phenomenological implications
of the generalized MFV ansatz for MFV scenarios defined both at the EW scale and at the GUT scale.

In particular, we want to address the question whether $\mathcal O(1)$ phases for the MFV
coefficients $c_i$, which are the only source of CPV in our setup, are phenomenologically allowed.

\section{EW scale MFV scenarios}
\label{EW_MFV}

The generalized MFV ansatz described in the previous section, where the $A_{i}$ terms are
assumed to be the only sources of CPV, implies a hierarchical structure for $\text{Im}\,A_{i}$.
In particular, it turns out that $\text{Im}\, A_t\gg \text{Im}\, A_{c,u}$, $\text{Im}\, A_b\gg\text{Im}\,A_{s,d}$ and $\text{Im}\, A_{\tau}\gg \text{Im}\, A_{\mu,e}$ (as $\text{Im}\,A_{i}$ scale with the cube of the fermion masses) and this leads to a natural suppression for the one loop SUSY contributions to the EDMs.

\begin{figure}[tb]
\includegraphics[width=0.235\textwidth]{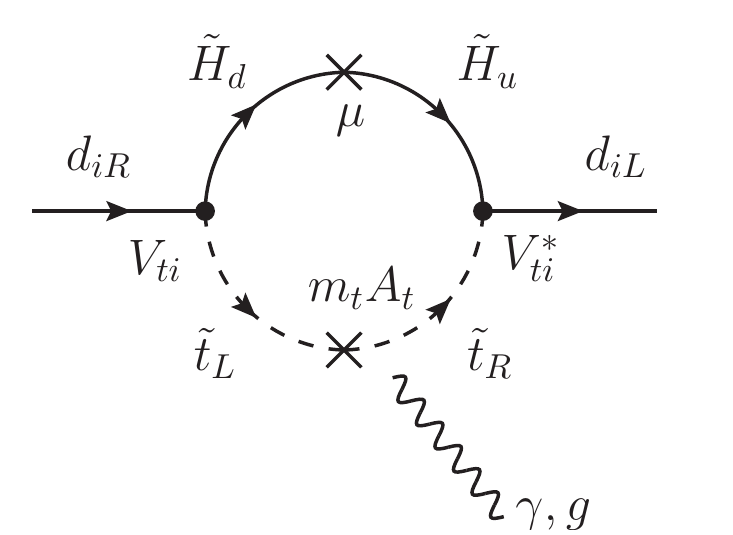}
\includegraphics[width=0.235\textwidth]{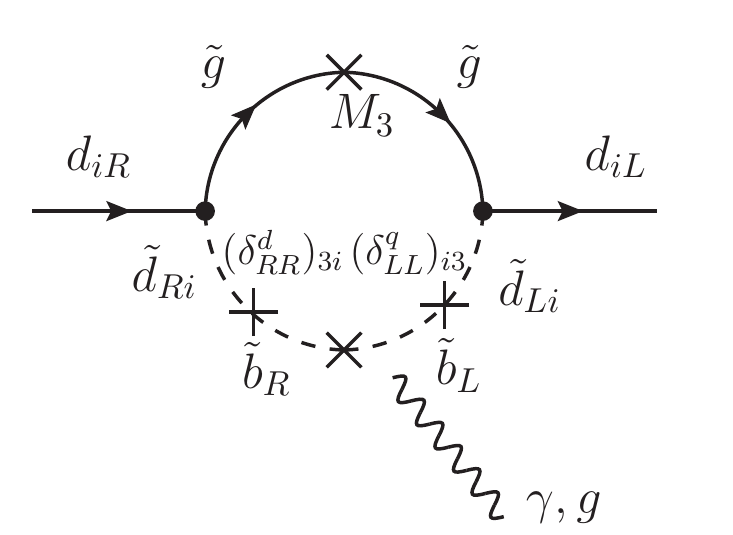}
\includegraphics[width=0.235\textwidth]{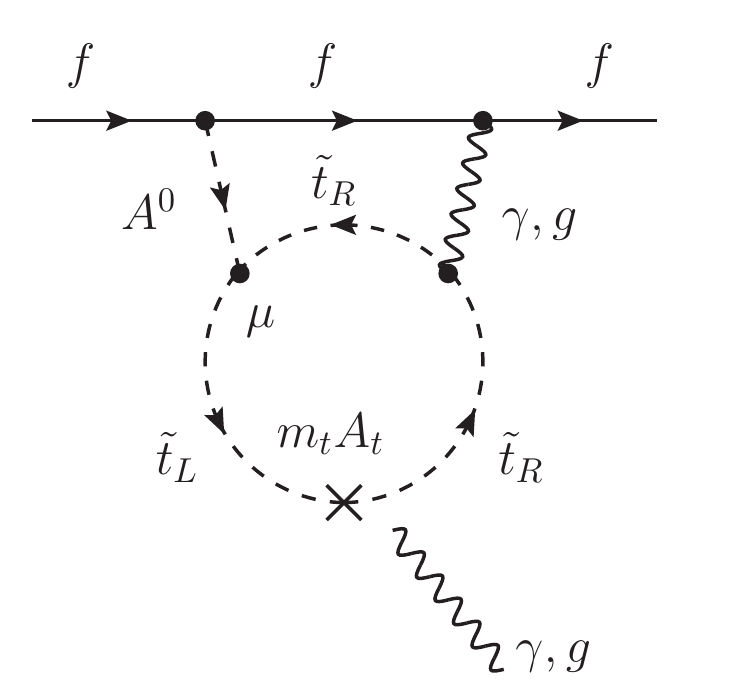}
\caption{Relevant SUSY contributions to the fermion EDMs in the EW scale MFV scenarios.
Upper Left: dominant one-loop contribution to the quark EDM generated by 
${\rm Im}(\mu A_{t})\neq 0$. Upper Right: flavor effect contributions to the quark
EDM mediated by the one-loop exchange of gluino/down-squarks. Lower: two-loop Barr-Zee
type diagram generating an EDM for quarks ($f=q$) and leptons ($f=\ell$) when
${\rm Im}(\mu A_{t})\neq 0$.}
\label{c78}
\end{figure}

Still, a potentially relevant one loop effect for the down quark EDMs, proportional to $\text{Im}A_t$, is induced by the stop exchange, as shown in the left-hand diagram of Fig.~\ref{c78}. It reads
\begin{eqnarray}
 \left\{ \frac{d_{d_i}}{e}\right\}_{{\tilde \chi}}
 \!\!\!\!\!\!~\simeq~\!\!\!\!
 -\frac{\alpha_2}{16\pi}\frac{5}{18}\frac{m_{d_i}}{m_{\tilde{q}}^4}
\frac{m_t^2}{m_W^2}{\rm Im}\left(A_t \mu\right)|V_{ti}|^2\tgb\,,
\label{charginoMFV}
\end{eqnarray}
leading to $d_n\sim 3\times 10^{-29}\,e$cm for maximum CPV phases, $m_{{\rm SUSY}}\sim 500\,\rm{GeV}$
and $t_{\beta}=10$, still far from $d^{\rm{exp}}_{n}\lesssim 10^{-26}\,e$cm.
The enhancement to $d_{d_i}$ induced by $\text{Im}A_t$ is compensated by the strong suppressing
factor $|V_{ti}|^2$.

A much more important effect is provided by the two loop Barr-Zee type diagram of Fig.~\ref{c78},
also involving only the third sfermion generation~\cite{pilaftsis}. These diagrams will generate
the electron EDM $d_e$ as well as the EDMs and chromo-EDMs for quarks. In particular, it turns out
that $d_e$ and the Mercury EDM $d_{\rm Hg}$ (as induced by the down-quark chromo-EDM) are the most
sensitive observables to this scenario. However, the theoretical estimation of $d_{\rm Hg}$ passes
through some nuclear calculations that unavoidably suffer from sizable uncertainties~\cite{pospelov}
hence, in the following, we focus on the predictions for $d_e$, to be conservative.

The induced electron EDM $d_e$ reads
\beq
  \frac{d_e}{e{\rm cm}}\simeq
  10^{-27}\left(\frac{500 {\rm GeV}}{m_{{\rm SUSY}}}\right)^{2}
  \left(\frac{t_{\beta}}{10}\right)\sin(\phi_{\mu}+\phi_A)\,,
  \label{barrzee_approx}
\eeq
where in Eq.~(\ref{barrzee_approx}) we have assumed $m_{{\rm SUSY}}=m_A$. Thus, if ${\cal O}(1)$ phases
are allowed, $d_e$ can reach the current experimental bound for $m_{{\rm SUSY}}\sim 500\rm{GeV}$ and $t_{\beta}=10$.

So far, we have not considered the contributions to the EDMs stemming from flavor effects~\cite{edm}. Indeed, the MFV flavor structures of Eqs.~(\ref{MFV:mQ})--(\ref{MFV:AD}) provide additional one loop ``flavored'' effects to the hadronic EDMs.

The off-diagonal terms of Eqs.~(\ref{MFV:mQ})--(\ref{MFV:AD}) can be conveniently
parameterized by means of the so-called MI parameters~\cite{MI} defined as usual as
\begin{align}
\delta^{L}_{ij} = (\mathbf{m}_{Q}^{2})_{ij}/\overline{m}_{Q}^{2} ~~\text{and}~~
\delta^{R}_{ij} = (\mathbf{m}_{D}^{2})_{ij}/\overline{m}_{D}^{2}
\end{align}
with $\overline{m}_{X}^{2}=\sqrt{(\mathbf{m}_{X}^{2})_{ii}(\mathbf{m}_{X}^{2})_{jj}}$ and $X=Q,L$.
Then, from Eqs.~(\ref{MFV:mQ})--(\ref{MFV:AD}), it follows that
\bea
\delta^{L}_{3i}&\simeq& (r_1 + c_1 y^2_b)\, V_{ti}\\
\delta^{R}_{3i}&\simeq& y_{b}y_{d_i}(r_4 + c_2 y^2_b)\, V_{ti}\,.
\label{MI_MFV}
\eea
One of the most important ``flavored'' effects to the hadronic EDMs arises from the gluino/squark
contribution shown in Fig.~\ref{c78}, leading to
\begin{eqnarray}
\left\{ \frac{d_{d_i}}{e}\right\}_{\tilde g}
\!\!\!\!~\simeq~\!\!\!\!
\frac{\alpha_s}{4\pi}
\frac{m_{\tilde g}}{m_{\tilde{q}}^2}
\frac{4}{135}{\rm Im}\left[\delta^{L}_{i3}\delta^{LR}_{33}\delta^{R}_{3i}\right]\,,
\label{gluinoEDM}
\end{eqnarray}
where $\delta^{LR}_{33} = m_b(A_b-\mu\tgb)/m_{\tilde{q}}^2$. The apparent bottom Yukawa enhancement
of Eq.~(\ref{gluinoEDM}), by means of $(\delta^d_{LR})_{33}\sim m_b$, is not effective within a SUSY
MFV scenario as the necessary $\delta^{R}$ MI turns out to be always proportional to light quark
Yukawas, see Eq.~(\ref{MI_MFV}).
In the most favorable situation, where we assume maximum CPV and $y_b\approx y_t \approx 1$, we find
$|\left\{d_{d_i}\right\}_{\tilde g}|\simeq 3\times 10^{-29}\,e$cm for $m_{{\rm SUSY}}=500\,\text{GeV}$
and $\mathcal O(1)$ parameters $|r_i|$, $|c_i|$.

Still, the two loop contributions of Fig.~\ref{c78} are largely dominant. The same conclusion holds
for all the other flavored effects to the hadronic EDMs, hence, we will not discuss them here.

Having discussed the dominant contributions to quark and lepton EDMs in the low-scale MFV setup,
we proceed now to assess its phenomenological viability in light of the experimental bounds on EDMs.
As an illustrative example, we choose a common SUSY mass $m_{SUSY}$ and consider separately the two
leading terms in the MFV expansion of $\mathbf{A}^{U}$, assuming purely imaginary coefficients to be
fair.

\begin{figure}[tb]
\includegraphics[width=0.45\textwidth]{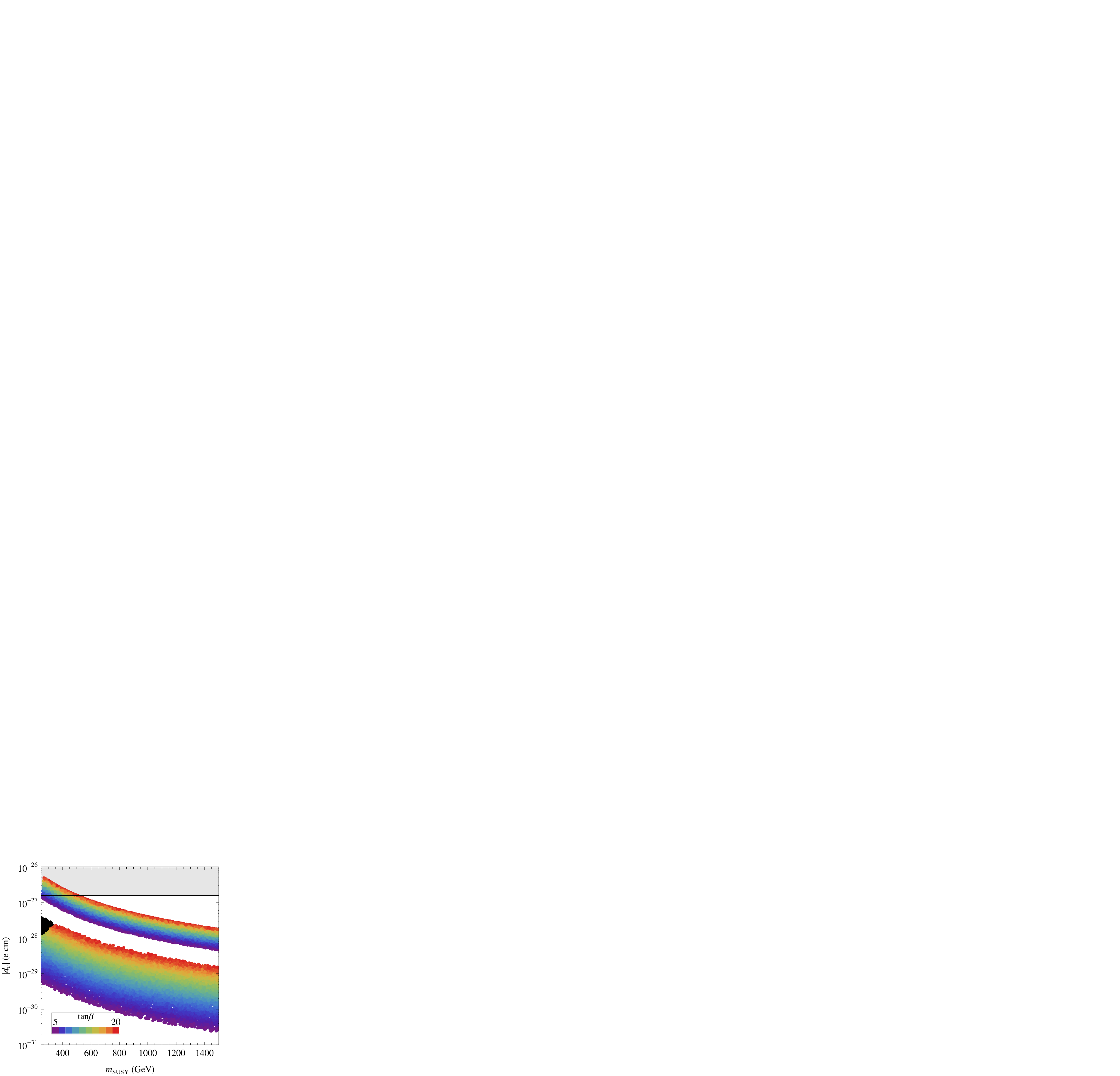}
\caption{Predictions for the electron EDM in a MFV framework defined at the EW scale.
The upper band corresponds to the scenario where 
$\mathbf{A}^{U}= A_{U}\mathbf{Y}_{u}(\mathbf{1}+c_4\mathbf{Y}_{u}^{\dagger}\mathbf{Y}_{u})$
with $c_4=i$,
while the lower band refers to the scenario where
$\mathbf{A}^{U}=A_{U}\mathbf{Y}_{u}(\mathbf{1}+c_3\mathbf{Y}_{d}^{\dagger}\mathbf{Y}_{d})$
with $c_3=i$.
In both cases, the black points are excluded by the constraints from B physics processes.}
\label{edm_mfv_lowscale}
\end{figure}

Consequently, in Fig.~\ref{edm_mfv_lowscale}, we show the predictions for the electron EDM $d_e$, as a
function of a common SUSY mass $m_{SUSY}$, arising within an EW scale MFV framework in these two cases:
\begin{enumerate}[i)]
\item $\mathbf{A}^{U}=A_{U}\mathbf{Y}_{u}(\mathbf{1}+c_4\mathbf{Y}_{u}^{\dagger}\mathbf{Y}_{u})$
\qquad with~$c_4=i$,
\item $\mathbf{A}^{U}=A_{U}\mathbf{Y}_{u}(\mathbf{1}+c_3\mathbf{Y}_{d}^{\dagger}\mathbf{Y}_{d})$
\qquad with~$c_3=i$.
\end{enumerate}
The most prominent feature of the two scenarios is their different scaling properties with $t_{\beta}$: in case i) $d_e\sim t_{\beta}$ while in case ii) $d_e\sim t^{3}_{\beta}$.
Moreover, the predictions for $d_e$ in the case ii) are suppressed compared to those of 
case i) by a factor of $y^2_b/y^2_t$.
Interestingly, Fig.~\ref{edm_mfv_lowscale} shows that $d_e$ is safely under control, but it
can reach experimentally visible levels, in both scenarios i) and ii) even for maximum CPV
phases and a light SUSY spectrum.

We conclude noting that, within an EW scale MFV scenario, the EDMs receive the dominant effects
at the two loop level while CPV effects in $B$-physics observables arise already at one loop~\cite{ABP}; hence large effects in $B$-physics can be still expected while being compatible with the EDM constraints. In particular, the phenomenology arising from the scenario discussed 
in this section is very similar to that discussed in Ref.~\cite{ABP}.
\section{GUT scale MFV scenarios}
\label{GUT_MFV}
In the previous section, we have assumed that the MFV expansion for the soft-breaking terms
of Eqs.~(\ref{MFV:mQ}),~(\ref{MFV:AD}) holds at the weak scale.

In contrast, in this section, we address the phenomenological implications for a MFV scenario defined at the high scale~\cite{runningMFV,colangelo_MFV}. In fact, even if we start with 
universal soft masses and proportional trilinear terms at the high-energy SUSY breaking scale 
$M_X$ (corresponding to setting all the coefficients $r_i$ and $c_i$ to zero) RGE effects do
not preserve such a universality. The MFV coefficients $r_i$ are RGE generated and their typical size is $(1/4\pi)^2\ln M_{X}^{2}/m_{{\rm SUSY}}^2$, so for sufficiently large values of $M_X$, 
the effect can be significant.

Moreover, as already discussed in Sec.~\ref{CPV_MFV}, it might be possible that the MFV flavor structure of the soft-sector can arise from an underlying flavor symmetry holding at some high energy scale.

In this respect, it seems quite natural to define a MFV scenario at the high scale.

As seen in Sec.~\ref{EW_MFV}, a remarkable virtue of a low-scale MFV scenario is its natural
solution to the SUSY CP problem by means of hierarchical $A$ terms.

However, generational hierarchies in the trilinear couplings are affected by RG effects since the
$A$ terms are not protected by the non-renormalization theorem. Therefore, even if these couplings
are assumed to vanish at the GUT scale, they can be regenerated through running effects.

This fact is particularly relevant for the impact of complex trilinears on quark or lepton EDMs.
For example, consider the RG equation for the up-squark trilinear; neglecting Yukawa couplings of
the two light generations and $U(1)$ gauge couplings, it reads
\begin{equation}
16 \pi^2 \frac{d}{dt} A_u= 6 A_t y_t^2 - 6 g_2^2 M_2 - \frac{32}{3} g_3^2 M_3\,,
\label{RGE:Au}
\end{equation}
where $t=\ln(\mu/\mu_0)$. The first term on the right-hand side of Eq.~(\ref{RGE:Au}) clearly
shows that, even if the gaugino mass terms are real, $A_u$ can receive a sizable imaginary
part if the stop trilinear $A_t$ is complex, with potentially dangerous impact on the one-loop contribution to the neutron EDM.

Approximate numerical expressions accounting for the low energy values of $A_u(m_Z)$ and $A_{t}(m_Z)$ as a function of the high energy input parameters, valid for low to intermediate $\tan\beta$,
are
\bea
A_u(m_Z) \!\!&\approx&\!\!
A_u(m_G)- 0.41 y_t^2 A_U + 0.03 y_b^2 A_D
\nonumber\\
&-&\!\!\left( 0.05 y_t^2 y_b^2 c_{3} + 0.11 y_t^4 c_{4}\right) A_U - 2.8m_{1/2},
\label{Au_fit}
\eea
\bea
A_t(m_Z) \!\!&\approx&\!\!
A_t(m_G)- 0.81 y_t^2 A_U
- 0.09 y_b^2 A_D
\nonumber\\
&+&\!\!\left( 0.04 y_t^2 y_b^2 c_{3} + 0.10 y_t^4 c_{4}\right) A_U
\nonumber\\
&-&\!\!\left( 0.03 y_t^2 y_b^2 c_{7} + 0.01 y_b^4 c_{8}\right) A_D - 2.2 m_{1/2}.
\label{At_fit}
\eea
where the Yukawa couplings are to be evaluated at the low scale and we have neglected terms of
$\mathcal O(y_i^6)$. Eq.~(\ref{Au_fit}) shows that, irrespective of $A_{u}(m_G)$, a sizable
contribution to $A_u(m_Z)$ from a complex $A_{t}(m_G)$ is unavoidable.

\begin{figure}[tb]
\includegraphics[width=0.45\textwidth]{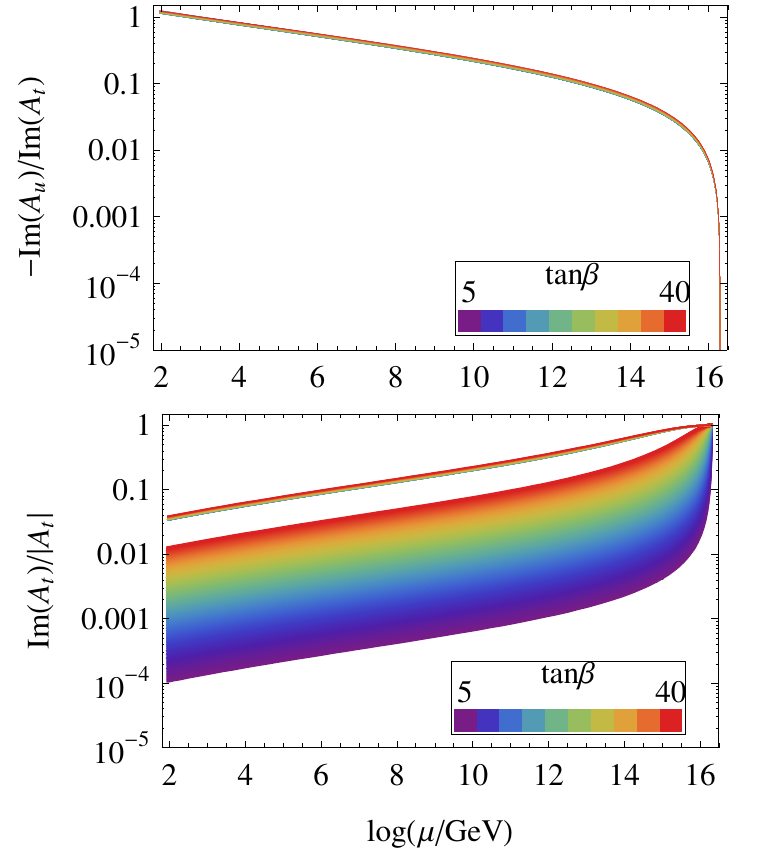}
\caption{Running of the trilinear terms in MFV scenarios defined at the GUT scale.
Upper plot: predictions for $\text{Im}A_{u}/\text{Im}A_{t}$ as a function of the renormalization
scale $\mu$ assuming the GUT scale boundary condition $\text{Im}A_{u}(m_G)=0$ and $A_t(m_G)\neq 0$.
Lower plot: predictions for $\text{Im}A_{t}/|A_{t}|$ as a function of the renormalization
scale $\mu$ assuming the GUT scale boundary condition 
$\mathbf{A}^{U} = c_4 A_{U}\mathbf{Y}_{u}\mathbf{Y}_{u}^{\dagger}\mathbf{Y}_{u}$ 
with $c_4=i$ (upper line) and
$\mathbf{A}^{U} = c_3 A_{U}\mathbf{Y}_{u}\mathbf{Y}_{d}^{\dagger}\mathbf{Y}_{d}$ 
with $c_3=i$ (lower band). $A_U=A_0=m_{1/2}$ was assumed for both plots.
}
\label{A_terms_highscale}
\end{figure}

This is well illustrated in the upper plot of Fig.~\ref{A_terms_highscale} where we show the
predictions for the ratio $\text{Im}A_{u}/\text{Im}A_{t}$ as a function of the renormalization scale
$\mu$ assuming the GUT scale boundary condition $\text{Im}A_{u}(m_G)=0$ and $\text{Im}A_t(m_G)\neq 0$.
Interestingly, the attained low energy values for $|\text{Im}A_{u}|$ and $|\text{Im}A_{t}|$
are very similar in spite of their very different values at the GUT scale, as is confirmed by
Eqs.~(\ref{Au_fit}), (\ref{At_fit}).

At the same time, huge RGE effects driven by the $SU(3)$ interactions strongly reduce the phase of
$A_u(m_Z)$ and $A_t(m_Z)$ (see Eqs.~(\ref{Au_fit}), (\ref{At_fit})), provided the gaugino masses are
real, as we assume.

This is illustrated in the lower plot of Fig.~\ref{A_terms_highscale}, where we show the predictions
for $\text{Im}A_{t}/|A_{t}|$ as a function of the renormalization scale $\mu$ assuming the GUT scale
boundary condition 
i) $\mathbf{A}^{U} =c_4 A_{U}\mathbf{Y}_{u}\mathbf{Y}_{u}^{\dagger}\mathbf{Y}_{u}$ setting $c_4=i$ 
(upper line), ii) $\mathbf{A}^{U}=c_3 A_{U}\mathbf{Y}_{u}\mathbf{Y}_{d}^{\dagger}\mathbf{Y}_{d}$ 
setting $c_3=i$ (lower band) and assuming $A_U=A_0=m_{1/2}$ in both cases.
We see that, even if we start with purely imaginary $A_t(m_G)$ at the GUT scale, such that
$\text{Im}A_{t}(m_G)/|A_{t}(m_G)|=1$, RGE effects reduce the phase of $A_t$ by more than one
order of magnitude in case i) and up to four orders of magnitude in case ii) depending on the
$\tan\beta$ value.

As a result, the attained values for $A_u(m_Z)$ and $A_t(m_Z)$ lie within an experimentally
allowed level for large regions of the parameter space even for $\mathcal{O}(1)$ phases, 
ameliorating significantly the SUSY CP problem.

Another even more dangerous CP violating contribution driven by the RGE effects regards the
$\mu$ term. To see this point explicitly, let's consider the one loop RGE for the $\mu$ parameter
\begin{eqnarray}
\frac{d\mu}{dt} = \frac{\mu}{16\pi^2}\left(-3g_2^2 + y_\tau^2 + 3y_b^2 + 3y_t^2\right)\,.
\label{RGE:mu}
\end{eqnarray}
As we can see, the phase of $\mu$ does not run and this is still true at the two loop level.
On the other hand, the RGE for the bilinear mass term $B$ is
\begin{eqnarray}
\frac{dB}{dt} = \frac{1}{8\pi^2}
\left( -3g_2^2\, M_2 + y_\tau^2 A_\tau + 3 y_b^2 A_b + 3 y_t^2 A_t
\right)\,,
\label{RGE:B}
\end{eqnarray}
then, in contrast to the $\mu$ term, the phase of the $B$ term is affected, through RGE effects,
by the phases of the $A$ terms. To have an idea of where we stand, it is useful to provide a
numerical solution to Eq.~(\ref{RGE:B}) as a function of the high scale parameters
\bea
B &\approx&
B_0 + \left( 0.15+0.60\tilde{t}^2\right)m_{1/2}
\nonumber\\
&-& 0.41 y_t^2 A_U - 0.42 y_b^2 A_D - 0.30 y_{\tau}^{2} A_E
\nonumber\\
&-& \left( 0.05 y_t^2 y_b^2 c_{3} + 0.11 y_t^4 c_{4}\right) A_U
\nonumber\\
&-& \left( 0.12 y_t^2 y_b^2 c_{7} + 0.05 y_b^4 c_{8} \right) A_D \,.
\eea
where $\tilde{t}=\tan\beta/50$, the Yukawa couplings are to be evaluated at the low scale and we have
again neglected terms of $\mathcal O(y_i^6)$.

Recall that, since the overall phase of the $\mu$ and $B\mu$ terms can be removed by a Peccei-Quinn 
transformation, only their overall phase is physical; moreover, the phase and absolute value of the
$B\mu$ term at the low scale is dictated by the EWSB conditions: in fact, in the basis where the
Higgs VEVs are real, these conditions require a real $B\mu$ term at the leading order~\footnote{Beyond
the leading order, the $B\mu$ term acquires a small imaginary part in the presence of CP violation in
the $\mu$ or $A$ terms, in order to compensate the CP-odd tadpole counterterms~\cite{Pilaftsis:1998pe,Pilaftsis:1999qt,Demir:1999hj}, while its real part is corrected 
by CP-even tadpoles.}. Thus, the condition that this relative phase vanishes at the high scale implies 
that the $\mu$ term must be complex.

\begin{figure}[tb]
\includegraphics[width=0.45\textwidth]{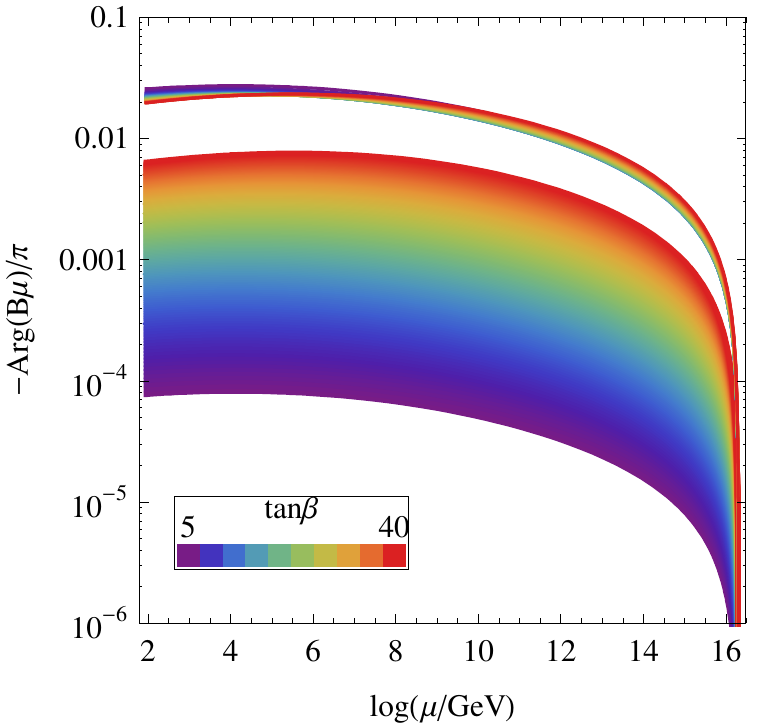}
\caption{Running of the phase $\phi_{\mu}+\phi_{B}$ in a MFV framework defined at the GUT scale
with respect to the renormalization scale assuming the GUT scale boundary condition
$\mathbf{A}^{U} =c_4 A_{U}\mathbf{Y}_{u}\mathbf{Y}_{u}^{\dagger}\mathbf{Y}_{u}$ with $c_4=i$,
while the lower band refers to the scenario where
$\mathbf{A}^{U} =c_3 A_{U}\mathbf{Y}_{u}\mathbf{Y}_{d}^{\dagger}\mathbf{Y}_{d}$ with $c_3=i$.
$B_0=A_U=A_0=m_{1/2}$ was assumed in both cases.
}
\label{B_term_highscale}
\end{figure}

This is shown in Fig.~\ref{B_term_highscale}, where we consider the dependence of the phase of $\phi_{\mu}+\phi_{B}$ on the renormalization scale assuming the GUT scale boundary condition
$\mathbf{A}^{U} =c_4 A_{U}\mathbf{Y}_{u}\mathbf{Y}_{u}^{\dagger}\mathbf{Y}_{u}$ with $c_4=i$
(upper line) and $\mathbf{A}^{U} =c_3 A_{U}\mathbf{Y}_{u}\mathbf{Y}_{d}^{\dagger}\mathbf{Y}_{d}$
with $c_3=i$ (lower band), and assuming $A_U=A_0=B_0=m_{1/2}$ for definiteness.

As we can see, the phase $\phi_{\mu}+\phi_{B}=0$ at the high scale as $\phi_{\mu}=0$ and $\phi_{B}=0$
singularly. However, at the low scale, $\phi_{\mu}+\phi_{B}\neq 0$ as the phase $\phi_{B}$ is
generated by RGE effects, in contrast to the phase $\phi_{\mu}$ that remains vanishing as it
does not run.

In principle, one can now impose by hand a real $\mu$ term at the EWSB scale (and hence at
all scales), but then the $B\mu$ term will be complex at the high scale; this is the approach that
is commonly assumed e.g. in the CMSSM. However, in our scenario, we assume that CP violation only
arises from the soft flavor-breaking terms of the MFV expansion, hence, the $B$ parameter at the 
high scale is assumed to be real.

\begin{figure}[tb]
\includegraphics[width=0.235\textwidth]{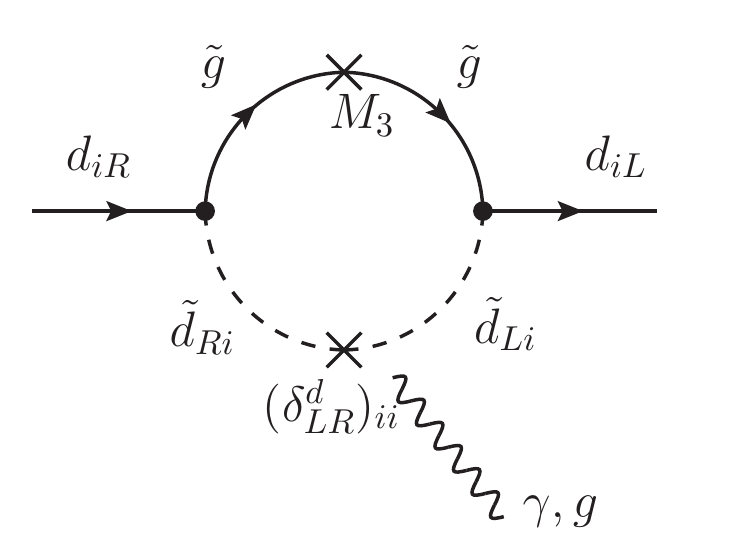}
\includegraphics[width=0.235\textwidth]{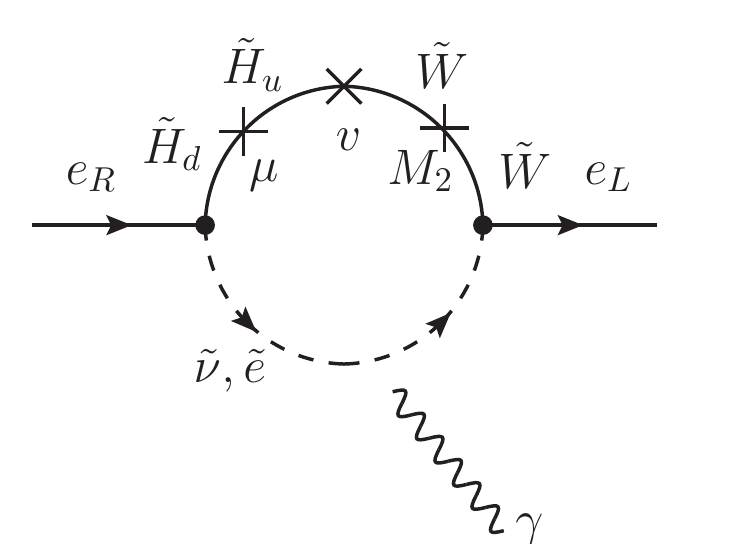}
\caption{Dominant one loop contributions to the EDMs of quarks (left-hand diagram)
and leptons (right-hand diagram) in a MFV framework defined at the GUT scale.}
\label{fig:edm2}
\end{figure}

Thus, if we start with a SUSY MFV scenario at the GUT scale, where CP violating sources are confined
to the third generation $A$ terms, at the low scale we unavoidably generate complex trilinears for
light generations and a complex $\mu$ term via RGE effects~\footnote{A related study within the 
CMSSM can be found in Ref.~\cite{Garisto:1996dj}}. As a result, the EDMs receive both one
and two loop contributions.
However, in our MFV scenario defined at the GUT scale, the dominant effects to the EDMs are by far
those induced by the one loop effects of Fig.~\ref{fig:edm2}, through the phase of the $\mu$ term.

After discussing the RGE effects leading to important contributions to quark and lepton EDMs
in the high-scale MFV setup, we proceed now to assess its phenomenological viability.

\begin{figure}[tb]
\includegraphics[width=0.45\textwidth]{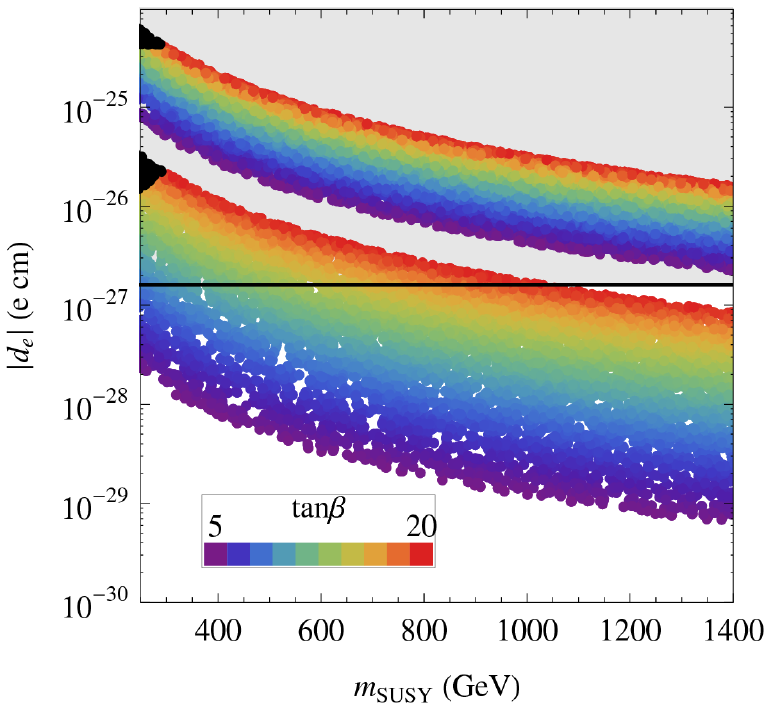}
\caption{Predictions for the electron EDM in a MFV framework defined at the GUT scale assuming
the boundary conditions
$\mathbf{A}^{U}=A_{U}\mathbf{Y}_{u}(\mathbf{1}+c_4\mathbf{Y}_{u}^{\dagger}\mathbf{Y}_{u})$
with $c_4=i$ (upper line) and
$\mathbf{A}^{U}=A_{U}\mathbf{Y}_{u}(\mathbf{1}+c_3\mathbf{Y}_{d}^{\dagger}\mathbf{Y}_{d})$
setting $c_3=i$ (lower band). 
$A_U=A_0=m_0=m_{1/2}\equiv m_\text{SUSY}$ was assumed in both scenarios.}
\label{edm_mfv_highscale}
\end{figure}

In Fig.~\ref{edm_mfv_highscale}, we show the predictions for the electron EDM in a MFV framework
defined at the GUT scale assuming the two cases i) and ii) for the trilinears $\mathbf{A}^{U}$
already discussed in the low-scale scenario; moreover, we set $A_U=A_0=m_0=m_{1/2}\equiv m_\text{SUSY}$.

As we can see, the scenario i) is ruled out for any $\tan\beta$ value up to a SUSY scale of order
$m_{{\rm SUSY}}\gtrsim 1.5$ TeV. On the contrary, the scenario ii) is still phenomenologically
allowed even for $m_{{\rm SUSY}}$ at the EW scale, provided $\tan\beta$ is moderate to small.

The above findings require some comments. In fact, from a phenomenological perspective, it seems
unlikely that the coefficient $c_4$ of Eq.~\ref{MFV:AU} can be an $\mathcal O(1)$ complex parameter,
as it would lead to the problems met in the above scenario i).
Let's try now to argue which could be the underlying theoretical motivation leading to a real $c_4$
making a comparison with the typical situations occurring in SUSY flavor models.

In these last cases, the flavor symmetries are spontaneously broken by the complex vacuum expectation
values of some ``flavon'' fields $\Phi$ and the hierarchical patterns for the Yukawa matrices are
explained in terms of suppressing factors $\left(\langle\Phi\rangle/M\right)^n$, as discussed in Sec.~\ref{CPV_MFV}. Clearly, given that the top Yukawa coupling is an order one parameter, it does
not require any suppressing factor and it is formally of the zeroth order in the
$\left(\langle\Phi\rangle/M\right)^n$ expansion.
%
Concerning the low energy phenomenology of the GUT MFV scenario discussed in this section,
we want to stress that the EDMs, arising at one loop level, are the most promising observables
and they generally prevent any visible effect in CPV $B$-physics observables. This is in contrast
with the EW MFV scenario where the EDM constraints were less stringent (as they arise at the
two loop level) and large $B$-physics signals, correlated with the predictions for the EDMs, were 
still allowed. However, the above features of the EW and GUT scenarios still cannot be considered
as an unambiguous tool to disentangle the two models.
In fact, also in the GUT MFV scenario, the main contributions to the EDMs can arise at the two-loop
level (see Fig.~\ref{c78}) in the context of hierarchical sfermions with light third and heavy first
generations~\cite{effective_susy}.
Should this be the case, the low-energy footprints of the EW and GUT MFV scenarios would turn out
to be indistinguishable and only a synergy of flavor data with the LHC data for the SUSY spectrum
could enable us to reconstruct the underlying scenario at work.
%
\section{Leptonic dipole moments: \boldmath $d_e$, $(g-2)_{\mu}$ and $\text{BR}(\ell_i\to\ell_j\gamma)$}
\label{sec:gm2_LFV}
In the following, we briefly discuss the correlations arising among dipole transitions in the leptonic sector~\cite{megletter}. In particular, we consider the electric dipole moment of the electron $d_e$,
the anomalous magnetic moment of the muon $a_{\mu}=(g-2)_{\mu}/2$ and the branching ratio of the lepton
flavor violating (LFV) decay $\mu\to e\gamma$ as these observables are highly complementary in shedding light on NP. In fact, while $a_{\mu}$ and $d_e$ are sensitive to the real and imaginary flavor diagonal dipole amplitude, respectively, $\text{BR}(\ell_i\to\ell_j\gamma)$ constrains the absolute value of off-diagonal dipole amplitudes.

Interestingly, most recent analyses of the muon $(g-2)$ point towards a $3\sigma$ discrepancy in 
the $10^{-9}$ range~\cite{g_2_th,passera_mh}: $\Delta a_{\mu}\!=\!a_{\mu}^{\rm exp}\!-\!a_{\mu}^{\rm SM}\approx(3\pm 1)\times 10^{-9}$. Hence, the question we intend to address now is which are the expected values for $d_e$ and $\text{BR}(\ell_i\to\ell_j\gamma)$ if we interpret the above discrepancy in terms of NP effects, in particular coming from SUSY.

As an illustrative case, if we consider the limit of a degenerate SUSY spectrum, the SUSY
contributions to $\Delta a_{\mu}$ and $d_e$ (as induced by flavor blind phases) read
\bea
\Delta a_{\mu}&\simeq&
\frac{\alpha_2}{4\pi}\,t_{\beta}\,\frac{5}{12}\frac{m^{2}_{\mu}}{m_{{\rm SUSY}}^{2}}\,,
\nonumber\\
\frac{d_e}{e} &\simeq&
\frac{\alpha_{2}}{4\pi}\,t_{\beta}\,\frac{5}{24}
\bigg(\frac{m_e}{m_{{\rm SUSY}}^2}\bigg)\sin\theta_\mu\,,
\label{edm_1loop}
\eea
leading to
\beq
\frac{|d_e|}{e\,\rm{cm}} \approx
10^{-27}\times\left(\frac{\Delta a_{\mu}}{3\times 10^{-9}}\right)\frac{|\theta_\mu|}{10^{-3}}\,.
\label{edm_gm2}
\eeq
The result of Eq.~(\ref{edm_gm2}) immediately leads to the conclusion that, as long as SUSY effects
account for the $(g-2)$ anomaly, the prediction for $d_e$ typically exceeds its experimental bound
$d_e\lesssim 10^{-27}$ unless $|\theta_\mu| \lesssim 10^{-3}$. An explanation for such a strong
suppression of $\theta_\mu$ can naturally arise within the general GUT MFV framework, as discussed
in the previous section. In fact, even assuming maximum CP violation in the high-scale trilinears
and setting the unknown MFV coefficient $c_3=1$, we have found that $|\theta_\mu|\gtrsim 10^{-4}$,
as shown in Fig.~\ref{B_term_highscale}.

Passing to ${\rm BR}(\ell_i\to \ell_j\gamma)$ and assuming again a degenerate SUSY spectrum, it is straightforward to find~\cite{Hisa1}
\beq
{\rm BR}(\mu\to e\gamma)\approx
2\times 10^{-12}
\left[\frac{\Delta a^{\rm SUSY}_{\mu}}{ 3 \times 10^{-9}}\right]^{2}
\bigg|\frac{\delta^{L}_{\mu e}}{10^{-4}}\bigg|^2\,.
\label{lfvgm2}
\eeq
where we have assumed that ${\rm BR}(\mu\to e\gamma)$ is generated only by the flavor structures
among left-handed sleptons, i.e. $\delta^{L}_{\mu e}$, as it happens in SUSY see-saw scenarios.

The main messages from the above relations is that within a GUT MFV SUSY scenario, with generalized
MFV ansatz, an explanation for the muon $(g-2)$ anomaly leads to predictions for $d_e$ that are close
to the current experimental upper bound $d_e\lesssim 10^{-27}e$~cm while ${\rm BR}(\mu\to e\gamma)$
typically lies within the expected MEG resolutions~\cite{meg} for values of $\delta^{L}_{\mu e}$
covering the predictions of many SUSY see-saw scenarios.
\section{Conclusions}
In this work we have addressed the SUSY CP problem in the framework of the MFV, where the SUSY flavor problem finds a natural solution. By contrast, the MFV principle does not solve the SUSY
CP problem as the MFV symmetry principle allows for the presence of new {\it flavor blind} CP-violating phases~\cite{colangelo_MFV,smithEDM} 
(see also~\cite{EllisCP,ABP,kaganMFV,feldmannNEW}).

Hence, the MFV ansatz has to be supplemented either by an extra assumption or by a mechanism accounting for a natural suppression of the {\it flavor blind} CPV phases.

In the light of these considerations, we have generalized the MFV ansatz accounting for a natural
solution of the SUSY CP problem.

We have assumed {\it flavor blindness}, i.e. universality of the soft masses and proportionality
of the trilinear terms to the Yukawas, when SUSY is broken.

In this limit, we have assumed CP conservation allowing for the breaking of CP only through the
MFV compatible terms breaking the {\it flavor blindness}.

That is, CP is preserved by the sector responsible for SUSY breaking, while it is broken in the flavor sector.

We have explored the phenomenological implications of this generalized MFV ansatz for MFV scenarios
defined both at the electroweak and at the GUT scales, pointing out the profound differences of
the two scenarios and their peculiar and testable predictions in low energy CP violating processes.


\textit{Acknowledgments:}
We thank A.~J.~Buras for useful discussions and comments on the manuscript.
This work has been supported in part by the Cluster of Excellence ``Origin and Structure
of the Universe'' and by the German Bundesministerium f{\"u}r Bildung und Forschung under
contract 05HT6WOA.



\begin{thebibliography}{999}


\bibitem{hfag}
  E.~Barberio {\it et al.}  [Heavy Flavor Averaging Group],
  arXiv:0808.1297 [hep-ex].


\bibitem{MFV}
  R.~S.~Chivukula and H.~Georgi, Phys. Lett. {\bf B188} (1987) 99;
  L.~J.~Hall and L.Randall, Phys. Rev. Lett. {\bf 65} (1990) 2939;
  A.~J.~Buras et al., Phys. Lett. {\bf B500} (2001) 161.

\bibitem{MFV_gen}
  G.~D'Ambrosio et al., Nucl. Phys. {\bf B645} (2002) 155.

\bibitem{colangelo_MFV}
  G.~Colangelo, E.~Nikolidakis and C.~Smith,
  Eur.\ Phys.\ J.\  C {\bf 59}, 75 (2009).

\bibitem{smithEDM}
  L.~Mercolli and C.~Smith,
  Nucl.\ Phys.\  B {\bf 817}, 1 (2009).

\bibitem{EllisCP}
  J.~R.~Ellis, J.~S.~Lee and A.~Pilaftsis,
  Phys.\ Rev.\  D {\bf 76}, 115011 (2007).


\bibitem{ABP}
W.~Altmannshofer, A.~J.~Buras and P.~Paradisi,
  Phys.\ Lett.\  B {\bf 669}, 239 (2008).


\bibitem{kaganMFV}
  A.~L.~Kagan et al.,
  arXiv:0903.1794 [hep-ph].

\bibitem{feldmannNEW}
  T.~Feldmann, M.~Jung and T.~Mannel,
  arXiv:0906.1523 [hep-ph].

\bibitem{smith_private}
We thank C.~Smith for drawing this point to our attention.


\bibitem{pospelov}
For a review of EDMs please see,
M.~Pospelov and A.~Ritz,
Annals Phys.\  {\bf 318}, 119 (2005) and therein references;
J.~R.~Ellis, J.~S.~Lee and A.~Pilaftsis,
  JHEP {\bf 0810}, 049 (2008).

\bibitem{oliveCP}
  K.~A.~Olive et al.,
  Phys.\ Rev.\  D {\bf 72}, 075001 (2005).

\bibitem{expedme}
  B.~C.~Regan et al.,
  Phys.\ Rev.\ Lett.\  {\bf 88}, 071805 (2002).

\bibitem{expedm}
  C.~A.~Baker {\it et al.},
  Phys.\ Rev.\ Lett.\  {\bf 97}, 131801 (2006).



\bibitem{abelian} M. Leurer, Y. Nir and N. Seiberg,
    {\sl Nucl. Phys.} {\bf B398} (1993) 319;
    {\sl Nucl. Phys.} {\bf B420} (1994) 468;
    L.E. Ibanez and G.G. Ross, {\sl Phys. Lett.} {\bf B332} (1994) 100;
    P. Binetruy and P. Ramond, {\sl Phys. Lett.} {\bf B350} (1995) 49;
    V. Jain and R. Shrock, {\sl Phys. Lett.} {\bf B352} (1995) 83;
    E. Dudas, S. Pokorski and C.A. Savoy, {\sl Phys. Lett.} {\bf B356} (1995) 45;
    P. Binetruy, S. Lavignac and P. Ramond, {\sl Nucl. Phys.} {\bf B477} (1996) 353.


\bibitem{nir_rattazzi}
  Y.~Nir and R.~Rattazzi,
  Phys.\ Lett.\  B {\bf 382}, 363 (1996).


\bibitem{nonabelian}
D.B. Kaplan and M. Schmaltz, {\sl Phys. Rev.} {\bf D49} (1994) 3741;
A. Pomarol and D. Tommasini, {\sl Nucl. Phys.} {\bf B466} (1996) 3;
L.J. Hall and H. Murayama, {\sl Phys. Rev. Lett.} {\bf 75} (1995) 3985;
C.D. Carone et al.,
{\sl Phys. Rev.} {\bf D54} (1996) 2328;
 R.~Barbieri et al, Phys.\ Lett.\  B {\bf 377}, 76 (1996);
 Nucl.\ Phys.\  B {\bf 493}, 3 (1997); Phys.\ Lett.\  B {\bf 401}, 47 (1997).


\bibitem{ross}
  G.~G.~Ross, L.~Velasco-Sevilla and O.~Vives,
  Nucl.\ Phys.\  B {\bf 692}, 50 (2004).

\bibitem{FN} C.D. Frogatt and H.B. Nielsen, {\sl Nucl. Phys.} {\bf B147} (1979) 277.


\bibitem{pilaftsis}
  D.~Chang, W.~Y.~Keung and A.~Pilaftsis,
  Phys.\ Rev.\ Lett.\  {\bf 82}, 900 (1999)
  [Erratum-ibid.\  {\bf 83}, 3972 (1999)].


\bibitem{edm}
  J.~Hisano, M.~Nagai and P.~Paradisi,
  Phys.\ Lett.\  B {\bf 642}, 510 (2006);
  Phys.\ Rev.\  D {\bf 78}, 075019 (2008);
  arXiv:0812.4283 [hep-ph].

\bibitem{MI}
L.~J.~Hall, V.~A.~Kostelecky and S.~Raby,
\npb{267}{1986}{415};
  F.~Gabbiani et al.,
  Nucl.\ Phys.\  B {\bf 477}, 321 (1996); 
  M.~Ciuchini et al.,
  Nucl.\ Phys.\  B {\bf 783} (2007) 112.


  \bibitem{runningMFV}
  P.~Paradisi, M.~Ratz, R.~Schieren and C.~Simonetto,
  Phys.\ Lett.\  B {\bf 668} (2008) 202.


\bibitem{Pilaftsis:1998pe}
  A.~Pilaftsis,
  Phys.\ Rev.\  D {\bf 58} (1998) 096010.

\bibitem{Pilaftsis:1999qt}
  A.~Pilaftsis and C.~E.~M.~Wagner,
  Nucl.\ Phys.\  B {\bf 553} (1999) 3.

\bibitem{Demir:1999hj}
  D.~A.~Demir,
  Phys.\ Rev.\  D {\bf 60} (1999) 055006.

 \bibitem{Garisto:1996dj}
   R.~Garisto and J.~D.~Wells,
   Phys.\ Rev.\  D {\bf 55} (1997) 1611.

\bibitem{effective_susy}
  A.~G.~Cohen et al., Phys.\ Lett. {\bf B388}, 588 (1996).


\bibitem{megletter}
  For a detailed analysis of this subject, please see
  J.~Hisano et al.,
  arXiv:0904.2080 [hep-ph].


\bibitem{g_2_th}
  M.~Passera,
  J.\ Phys.\ G {\bf 31} (2005) R75;
  Nucl.\ Phys.\ Proc.\ Suppl.\  {\bf 155} (2006) 365;
  M.~Davier,
  Nucl.\ Phys.\ Proc.\ Suppl.\  {\bf 169}, 288 (2007);
  K.~Hagiwara et al.,
  Phys.\ Lett.\  B {\bf 649}, 173 (2007).

\bibitem{passera_mh}
  M.~Passera, W.~J.~Marciano and A.~Sirlin,
  Phys.\ Rev.\  D {\bf 78}, 013009 (2008).


\bibitem{Hisa1}
  J.~Hisano and K.~Tobe,
  Phys.\ Lett.\ B {\bf 510}, 197 (2001);
  G.~Isidori et al.,
  Phys.\ Rev.\  D {\bf 75} (2007) 115019.

\bibitem{meg}
Talk given by Marco Grassi, Les Rencontres de Physique de la Vallee D'Aoste,
La Thuile, Aosta Valley, Italy 
(March 1-7, 2009). Presentatation is placed on
http://agenda.infn.it/conferenceDisplay.py?confId=930.

\end{thebibliography}
\end{document}